\documentclass[a4paper]{jpconf}
\usepackage{graphicx}
\usepackage{float}
\usepackage{psfrag}
\begin{document}

\title{Optimization of the half wave plate configuration for the LSPE-SWIPE experiment}

\author{A Buzzelli$^{1,4}$, G de Gasperis$^{3,4}$, P de Bernardis$^{1,2}$, S Masi$^{1,2}$ and \\N Vittorio$^{3,4}$}

\address{$^{1}$ Dipartimento di Fisica, ``Sapienza'' Universit\`a di Roma, piazzale Aldo Moro 5, I-00185, Roma, Italy}
\address{$^{2}$ Sezione INFN Roma 1, piazzale Aldo Moro 5, I-00185, Roma, Italy}
\address{$^{3}$ Dipartimento di Fisica, Universit\`a di Roma ``Tor Vergata'', via della Ricerca Scientifica 1, I-00133, Roma, Italy}
\address{$^{4}$ Sezione INFN Roma 2, via della Ricerca Scientifica 1, I-00133, Roma, Italy}

\ead{alessandro.buzzelli@roma2.infn.it}

\begin{abstract}
The search for the $B$-mode polarization of Cosmic Microwave Background (CMB) is the new frontier of observational 
Cosmology. A $B$-mode detection would give an ultimate confirmation to the existence of a primordial Gravitational 
Wave (GW) background as predicted in the inflationary scenario. Several experiments have been designed or planned 
to observe $B$-modes. In this work we focus on the forthcoming Large Scale Polarization Explorer (LSPE) experiment, 
that will be devoted to the accurate measurement of CMB polarization at large angular scales. LSPE consists of a  
balloon-borne bolometric instrument, the Short Wavelength Instrument for the Polarization Explorer (SWIPE), and a 
ground-based coherent polarimeter array, the STRatospheric Italian Polarimeter (STRIP). SWIPE will employ a rotating 
Half Wave Plate (HWP) polarization modulator to mitigate the systematic effects due to instrumental non-idealities. 
We present here preliminary forecasts aimed at optimizing the HWP configuration.
 
\end{abstract}

\section{Introduction}

The Cosmic Microwave Background (CMB) is the oldest light in the universe, freely streaming since the cosmic 
recombination epoch, characterized by a nearly isotropic black body spectrum with temperature anisotropies of 
only few parts out of $10^5$. The CMB has represented the most important tool to constrain the initial stage, 
evolution and content of the universe. In fact, the CMB anisotropy pattern has provided tight constraints on the 
cosmological parameters, such as the curvature and the expansion of the Universe and the relative abundances of 
baryons, Dark Matter and Dark Energy \cite{Pla}. However, to break the degeneracies among some parameters, an 
accurate analysis of the CMB (linear) polarization pattern is required.  
CMB polarization is generated at last scattering surface by Thomson scattering of quadrupolar anisotropies. 
A complete description of the polarization field can be achieved with the Stokes parameters $Q$ and $U$, 
as $I$ describes only temperature and $V=0$ as Thomson scattering induces no circular polarization. 
Since the polarization field can be encoded in a 2-rank symmetric trace-free tensor, we can decompose it into 
a gradient ($E$-modes) and a curl ($B$-modes) component. $E$-modes have been widely observed, while $B$-modes 
are still buried into instrumental noise and foreground contamination. Primordial scalar (i.e. density) perturbations 
produce only $E$-modes, while tensor perturbations produce both $E$ and $B$-modes. Therefore, a detection of primordial 
$B$-modes would provide a definitive confirmation to the existence of a stochastic Gravitational Wave (GW) background, 
as predicted in the inflationary paradigm. 
Several experiments have been designed or planned with the ultimate goal of detecting $B$-mode polarization, either 
from ground, balloons or space. In this work, we focus on the forthcoming LSPE mission that will be devoted to the 
accurate measurement of CMB polarization at large angular scales \cite{Aio}. LSPE will consist of a high frequency 
balloon-borne bolometric polarimeter, SWIPE \cite{deB}, and of a low frequency ground-based coherent polarimeter, 
STRIP \cite{Ber}. The peculiarity of SWIPE will be the presence of a rotating half wave plate (HWP) polarization 
modulator to mitigate the instrumental systematics.
The work is organized as follows: in Section 1 we  briefly overview the CMB $B$-mode polarization; in 
Section 2 we introduce the scientific case of the LSPE experiment; in Section 3 we present preliminary simulations 
aimed at optimizing the SWIPE HWP configuration; finally, we draw our conclusions in Section 4.

\section{CMB B-mode polarization}

The amplitude of $B$-modes is usually parametrized by the “tensor-to-scalar ratio”, $r$, that expresses the relative 
amplitude of tensor and scalar primordial perturbations. The satellite mission Planck, together with the ground based 
experiments BICEP2 and Keck Array, set the most stringent upper limit to date of $r < 0.07$ at 95\% C.L. \cite{BIC}, but we do not 
have precise constraints on its lower limit. However, theoretical arguments suggest $r > 10^{-3}$ \cite{Kam}.
The LSPE experiment is dedicated to observe CMB polarization at large angular scales, where most of the primordial 
$B$-mode information is encoded, with the primary aim to constrain $r$ down to around 0.02. This tight contraint will 
allow to rule out several inflationary scenarios, thanks to the strict relation between $r$ and the inflation energy 
scale: $E = 3.3 \times 10^{16} r^{1/4} \;GeV$.
Inflation is a stage of exponential expansion in the very early Universe that lasted for a fraction of second. 
This theory was introduced to solve the “classical” puzzles of the Hot Big Bang scenario, such as the horizon, 
the flatness and the monopole problems. From the Friedmann equations, a period of accelerated expansion implies 
an exponential growth for the scalar factor $a$, i.e. $a(t) \propto e^{Ht}$, where $H$ is the Hubble parameter, 
and an unusual relation for the equation of state, $p = w\rho$ with $w < −1/3$, where $p$ and $\rho$ indicate Universe 
mean pressure and density, respectively. Moreover, inflation provides a mechanism for the production of perturbations 
in the early Universe, both scalar and tensor, from primordial quantum fluctuations. According to the current 
understanding, cosmic structures have been originated by gravitational instability of density (scalar) primordial
perturbations, which are ``adiabatic'' 
and form a random Gaussian field with a nearly scale-invariant spectrum. Imprints of scalar perturbations are in 
both temperature and $E$-mode polarization patterns. Tensor perturbations are associated to a stochastic GW background, 
which impressed specific traces in the temperature, $E$-mode and, unequivocally, $B$-mode patterns. For this reason, 
a $B$-mode detection would represent a priceless breakthrough in Cosmology and fundamental physics. Moreover, $B$-mode 
polarization can shed light on some critical aspects of cosmic reionization, eventually associated to the first cosmic 
structure formation, which is expected to move part of $E$-modes into $B$-modes at very large angular scales.
Different inflationary scenarios do exist. The canonical single-field slow-roll model is described by only two 
parameters, $\epsilon$ and $\eta$ \cite{Kam}. The spectral index of the power spectrum of scalar perturbations is defined 
as $n_s = 2\eta - 6\epsilon +1$: 
inflation predicts a value of $n_s$ slightly less than 1. The most recent constraint by Planck 
is $n_s = 0.968 \pm 0.006$ \cite{Pla}. The tensor spectral index is given by $n_t = -2\epsilon$, and it is expected 
to be negative. 
CMB temperature ($T$) and polarization ($E$,$B$) fields can be described in terms of their auto ($TT$,$EE$,$BB$) 
and cross ($TE$,$TB$,$EB$) angular power spectra $C_l$, representing the power in fluctuations at each angular scale,
which is expressed in terms of the multipole moment $\ell$ in harmonic space. According to the standard cosmological 
model, $C_l^{TB} = C_l^{EB} = 0$ due to the parity properties of the CMB physics. Recent estimates of $C_l^{TT}$, 
$C_l^{TE}$  and $C_l^{EE}$ can be found in [1]. Notice that the prediction of Gaussian primordial fluctuations implies 
that also CMB fluctuations are expected to be Gaussian: this means that the power spectra fully describe the statistical 
anisotropy pattern. 
The spectrum $C_l^{BB}$ is characterized by a peak at $\ell \leq 10$ (reionization bump), due to the conversion 
from $E$-modes to $B$-modes occurred at redshift $z \sim  8$, and by a peak at $\ell \sim 80$ (recombination bump), 
that is the imprint of the inflationary GW background. 
We only mention here three important effects that may introduce a spurious (i.e. non primordial) B-mode component: 
systematic effects, Gravitational Lensing (GL) and Galactic foreground contamination. Systematic effects are due to 
instrumental non-idealities or data analysis artifacts. GL is caused by large scale mass inhomogeneities between us 
and the last scattering surface and its effect is dominant at small angular scales. Polarized foreground contamination 
is mainly due to synchrotron radiation, below $40\;GHz$, and thermal dust emission, above $90\;GHz$. These effects must 
be carefully controlled and possibily suppressed as they may introduce a B-mode component in general larger than the 
primordial B-modes, degrading thus the performance of any experiment.

\section{The LSPE experiment}

\begin{figure}[H]
\includegraphics[scale=0.28]{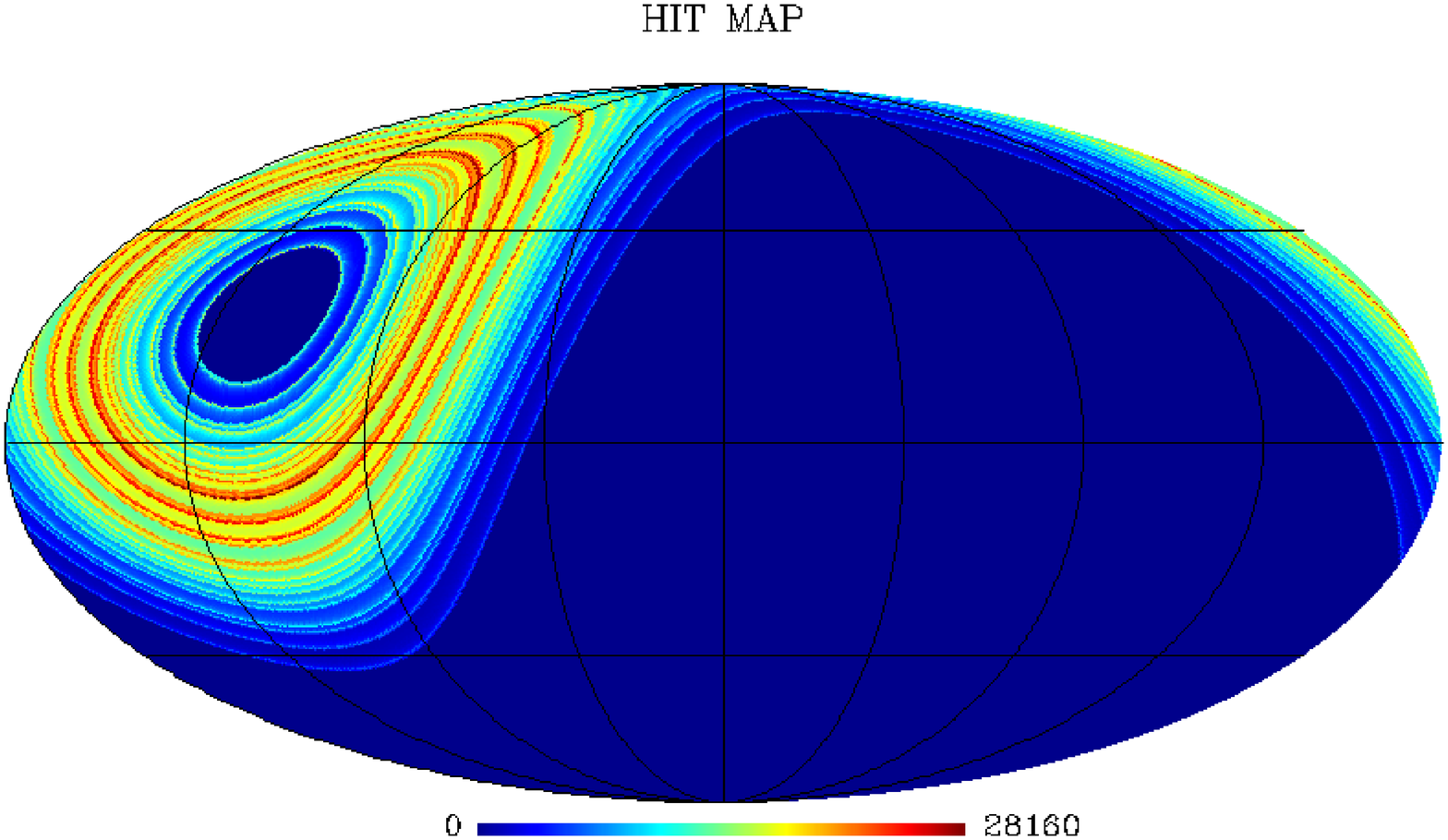}
\includegraphics[scale=0.28]{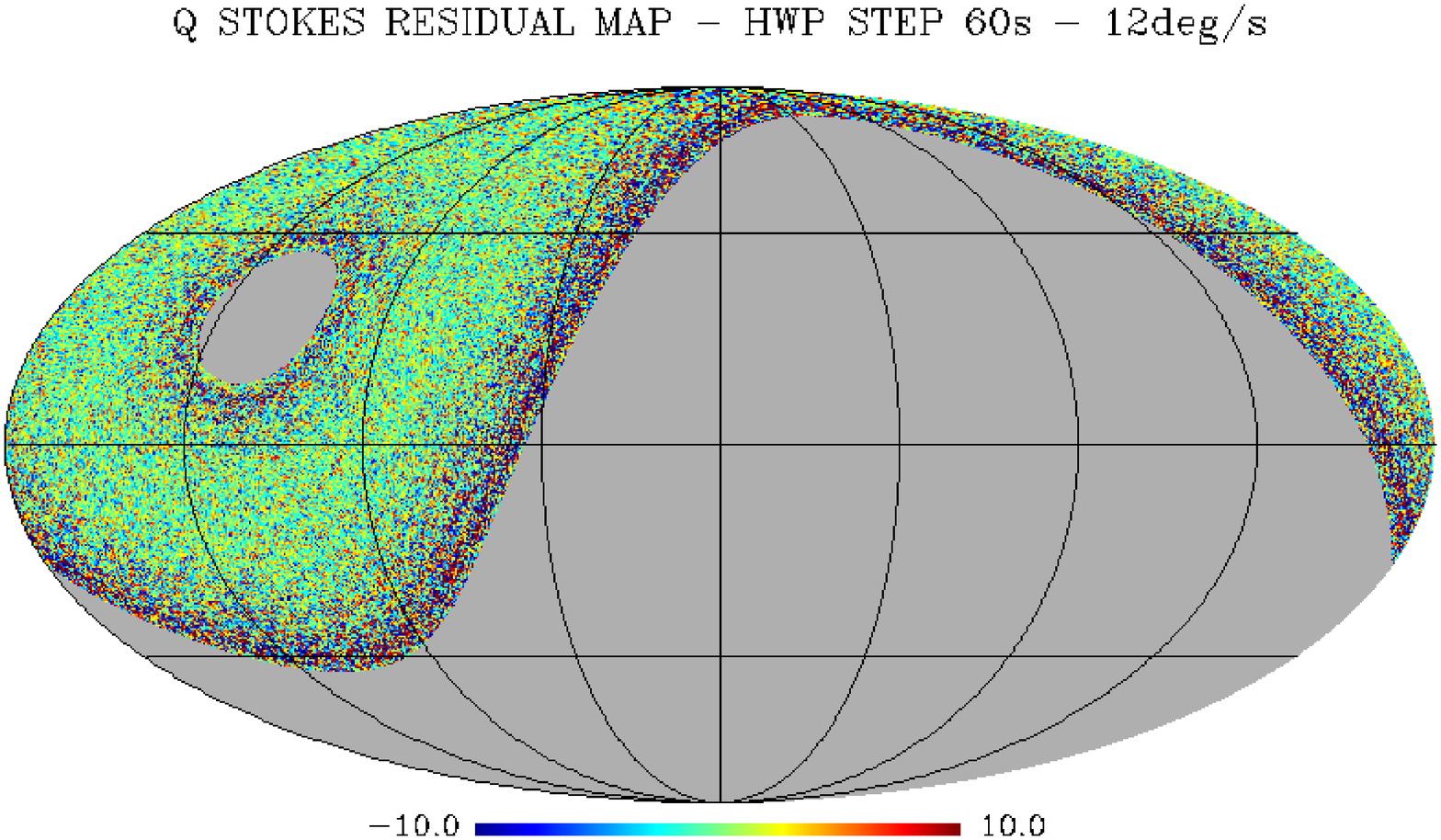}
\caption{Maps from LSPE-SWIPE simulations. On the left: $hitmap$, i.e. the pixel-weighted coverage map. 
On the right: residual map (in $\mu K$), i.e. the difference between the output optimal map and the input map (where also the 
noise has been added) for the $Q$ Stokes parameter. See the text for the details of the simulations.}
\label{maps}
\end{figure}

LSPE is a next-generation CMB experiment \cite{Aio}, devoted to the observation of large scale polarization, with the 
primary aim of constraining the B-mode signal due to reionization (reionization bump) and the rising part of the 
signature of primordial tensor perturbations (recombination bump). LSPE will observe a large fraction of the northern 
sky (around the 30\% of the celestial sphere) with a coarse angular resolution of about 1.5 degrees FWHM. LSPE will 
consist of two instruments: a balloon-based array of bolometric polarimeters, SWIPE, which will map the sky in three 
frequency bands centered at 140, 220 and $240\;GHz$ \cite{deB}, and a ground-based array of coherent polarimeters, 
STRIP, which will survey the same sky region in two frequency bands centered at 43 and $90\;GHz$ \cite{Ber}. 
This work is specific to the SWIPE instrument. SWIPE will be likely launched in winter 2018-2019 and will operate 
for around 15 days during the Arctic night (at latitude around $78\;N$), to exploit optimal observation conditions. 
SWIPE will scan the sky by spinning around the local vertical, while keeping the telescope elevation constant for 
long periods, in the range 35 to $55\;deg$. According to the current baseline, the azimuth telescope scan-speed will 
be set around 2 rpm, i.e. $12\;deg/s$. SWIPE will host a HWP polarization modulator as first optical element, followed 
by a $50\;cm$ aperture refractive telescope, a beamsplitting polarizer, and finally two symmetric orthogonally-placed 
focal planes with an overall number of 110 detectors per frequency. The nominal HWP configuration consists of a 
step-and-integrate mode, repeatedly scanning the range $0-78.75\;deg$ at a rate of $11.25/min$.
We generate sky-maps and angular power spectra from simulated observations in order to forecast the performances of 
the SWIPE instrument.
Our flight simulator provides the coordinates (right ascension and celestial declination) and the polarization angle 
based on the SWIPE scanning strategy. Hence, we produce power spectra from the CAMB software \cite{Lew} according to the 
last release of cosmological parameters from Planck \cite{Pla} assuming a tensor-to-scalar ratio $r=0.09$, that we use to
generate a full sky map with the Synfast facility of the HEALPix package \cite{Gor}. 
Finally, we scan this ``input'' map to produce the corresponding Time Ordered Data (TOD). 
We simulate the noise having a spectrum with a constant plateau at high frequencies with 
amplitude 15$\;\mu K s^{1/2}$, and a correlated $1/f^{\alpha}$ part at low frequencies with $\alpha=2$. Moreover, we 
include a cross-correlated noise component equally shared by all the detectors of the form $1/f^2$ at low frequencies. 
The knee frequency is set to $0.1\;Hz$. We limit our simulation to 18 detectors, arranged in 
three triples sparcely located in each focal plane, and to 5 days of operation with the telescope elevation 
switching 5$\;deg$ every day from 35 to 55$\;deg$. In our simulations we focus on the $140\;GHz$ band, that will be the 
main SWIPE CMB channel, while observations at 220 and $240\;GHz$ will be devoted to monitor the polarized thermal dust 
emission from the Galaxy. The maps are generated according to the nominal values of the telescope scan-speed and HWP 
(stepping) configuration. The angular resolution is set to 1.5 degree FWHM with a HEALPix $N_{side}=128$.
To produce the maps from the TODs we use the ROMA MPI-paralled code, an optimal map-making algorithm based on a generalized least 
squared approach \cite{deG}. The code has been extended to account for all the off-diagonal terms of the noise 
covariance matrix, that correspond to the cross-correlated noise among the detectors \cite{deG2}. Hence, we estimate 
the power spectra by a pseudo-$C_l$ approach, also knwon as the MASTER method \cite{Hiv}.
In Fig.~\ref{maps} we show the $hitmap$ (i.e. the pixel-weighted coverage map) and the residual map (i.e. the 
difference between the output optimal map and the input map, where also the noise has been added) for the $Q$ 
Stokes parameter. In Fig.~2 we show the average BB power spectrum estimated from 50 noise-only, signal-only and 
signal plus noise simulated maps (the error bars are given by the dispersion of the simulations). 

\begin{figure}[H]
\begin{center}
\psfrag{Multipole l}[c][][3]{Multipole $\ell$}
\psfrag{l(l+1)/(2pi) ClBB }[c][][3]{$\ell(\ell+1) C_l^{BB} /(2\pi$)}
\includegraphics[angle=90,scale=0.30]{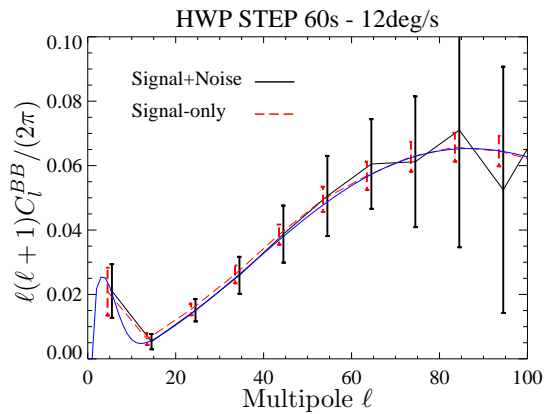}
\caption{BB angular power spectra from LSPE-SWIPE simulations. Average spectrum and associated error bars 
estimated from signal plus noise simulations (in solid black line) and from signal-only simulations (in dashed red line). The input power spectrum, 
corresponding to $r=0.09$, is shown for comparison in blue. Note the reionization bump at $\ell \leq 10$, and the recombination bump
at $\ell \sim 80$. See the text for the details of the simulations.}
\end{center}
\label{BB}
\end{figure}

\section{Optimization of the polarization modulation strategy}

In this Section we present preliminary forecasts on the optimization of the polarization modulation strategy for the 
SWIPE instrument. Wave plates are generally characterized by two orthogonal axes with different refractive indexes 
(the fast and the slow axis). A HWP is the particular case with a retardation of $\pi$ between the phases of the two 
orthogonal components of the incident radiation electric field. To be effective, HWPs must be rotated during the 
observation, either in stepping or (continuously) spinning mode. The incident polarization is thus modulated at 
4 $f_r$, where $f_r$ is the rotation frequency.  The use of a HWP as polarization modulator represents a powerful tool 
to minimize most of the instrumental systematic effects, that would otherwise severely degrade the experimental 
performance. A HWP allows to: i) effectively mitigate beam, calibration and other intrumental systematics, as a HWP 
design does not need to difference power 
from orthogonal polarization sensitive detectors, and additionally allows not to rotate the whole instrument; 
ii) reject the $1/f$ noise at the hardware level, as it moves the polarization signal to high frequencies allowing 
thus to sample the Stokes parameters $Q$ and $U$ in the white noise regime; iii) achieve a better angle coverage 
uniformity.

Here, we focus on two specific benefits provided by the HWP: the shift of the polarization signal to high frequencies 
and the improvement in the pixel angle coverage. The first effect allows to preserve cosmological information when a 
low-frequency cut of the data stream is performed (common practice to filter out the $1/f$ noise). The second effect 
helps the sky-map reconstruction since each pixel is observed from more directions. We investigate a HWP stepping 
every 1, 60 and $3600\;s$ and spinning at 5, 2 and $0.5\;Hz$, keeping fixed the telescope azimuth scan-speed at the 
nominal value of 2 rpm. In Fig.~3 we show the ``periodograms'', i.e. the plots of temperature and polarization 
intensity power as function of frequency. We notice that a stepping HWP is not effective in modulating the polarization 
to higher frequencies; on the contrary, a (fast) spinning HWP 
is extremely powerful in moving the polarization away from the low-frequency part dominated by the $1/f$ noise (for instance, 
a HWP spinning at $5\;Hz$ modulates the polarization signal in a narrow band around $20\;Hz$).

\begin{figure}[H]
\begin{center}
\psfrag{Power [uK^2/Hz]}[c][][3]{\normalsize{Power $[\mu K^2/Hz]$}}
\psfrag{Freq [Hz]}[c][][3]{\normalsize{Frequency $[Hz]$}}
\includegraphics[angle=90,width=0.32\textwidth]{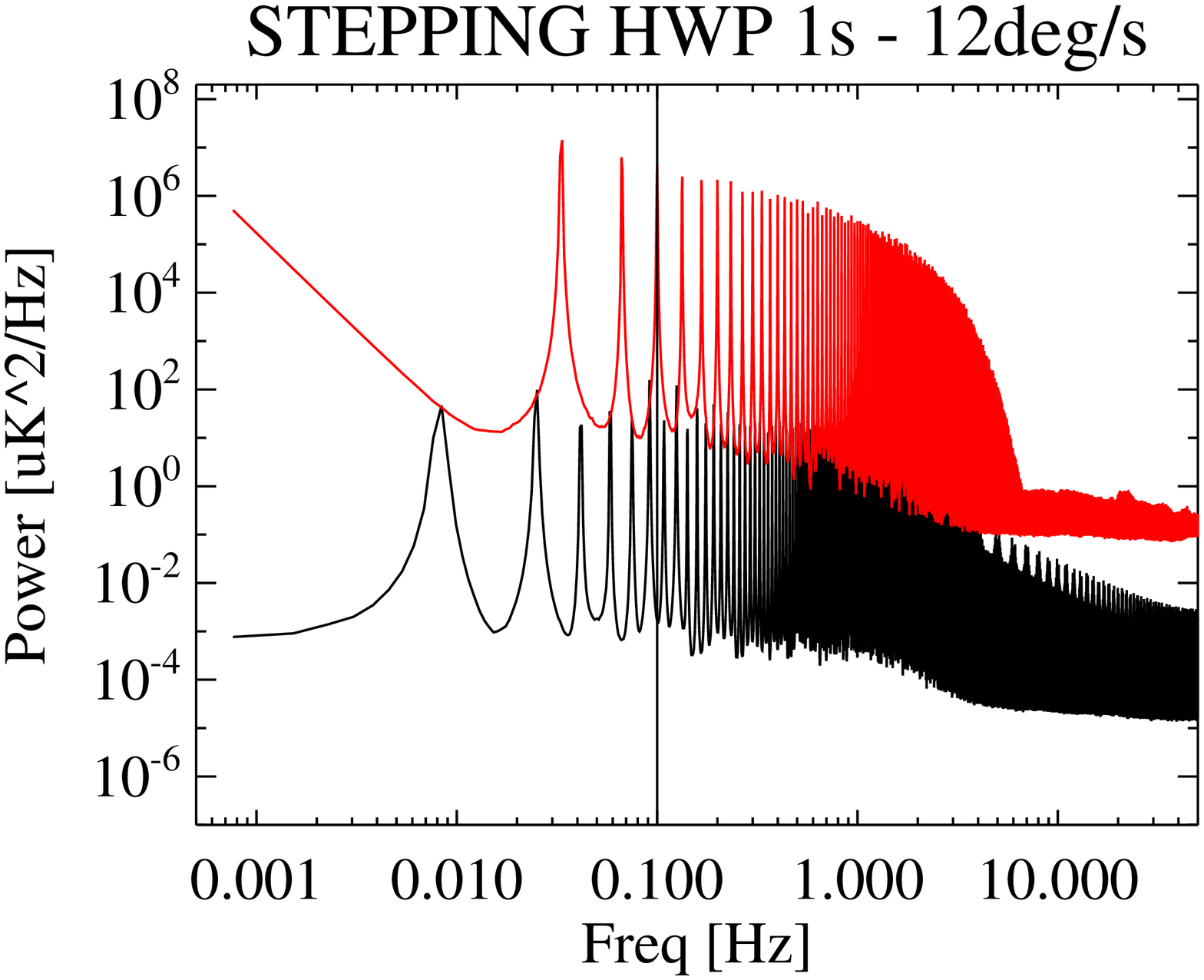}
\includegraphics[angle=90,width=0.32\textwidth]{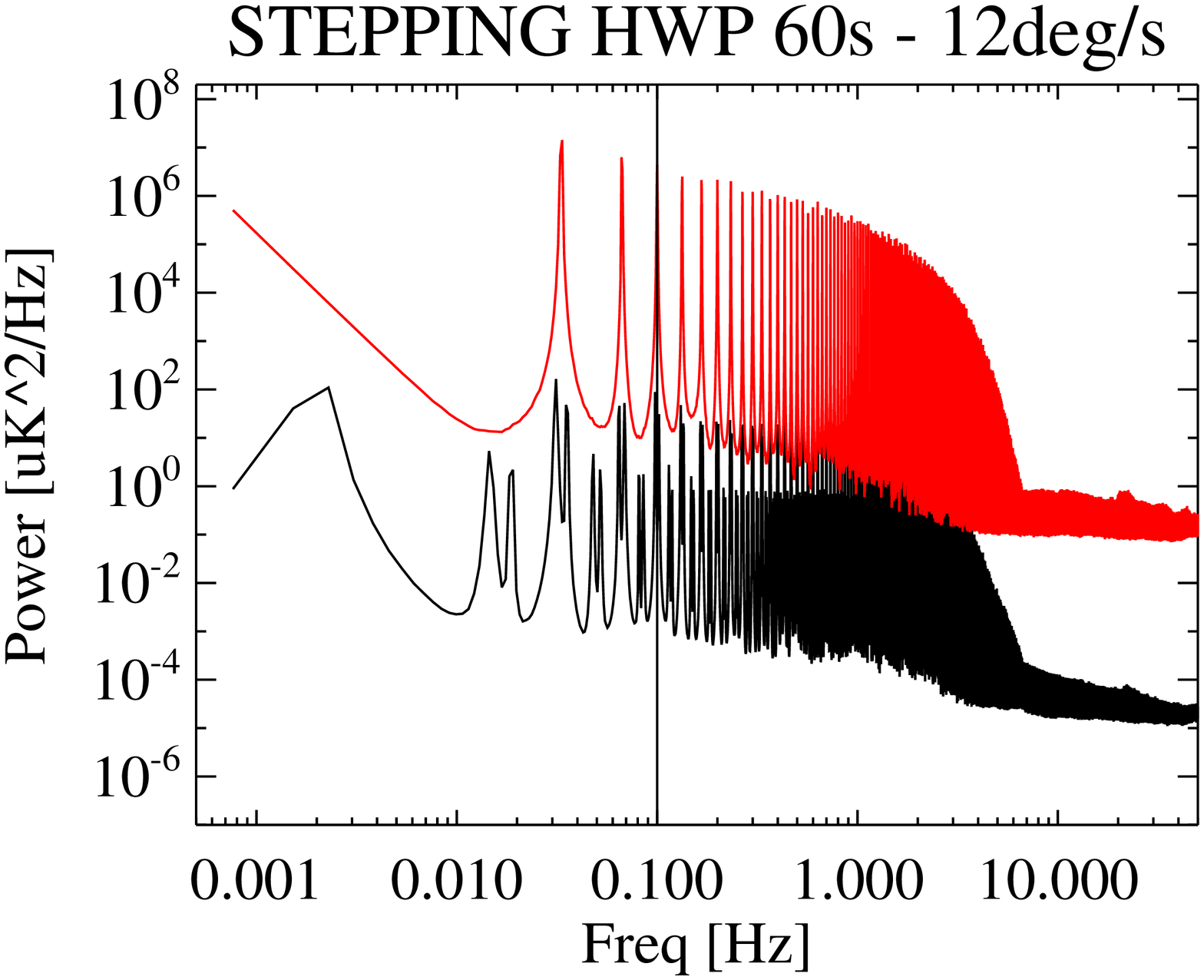}
\includegraphics[angle=90,width=0.32\textwidth]{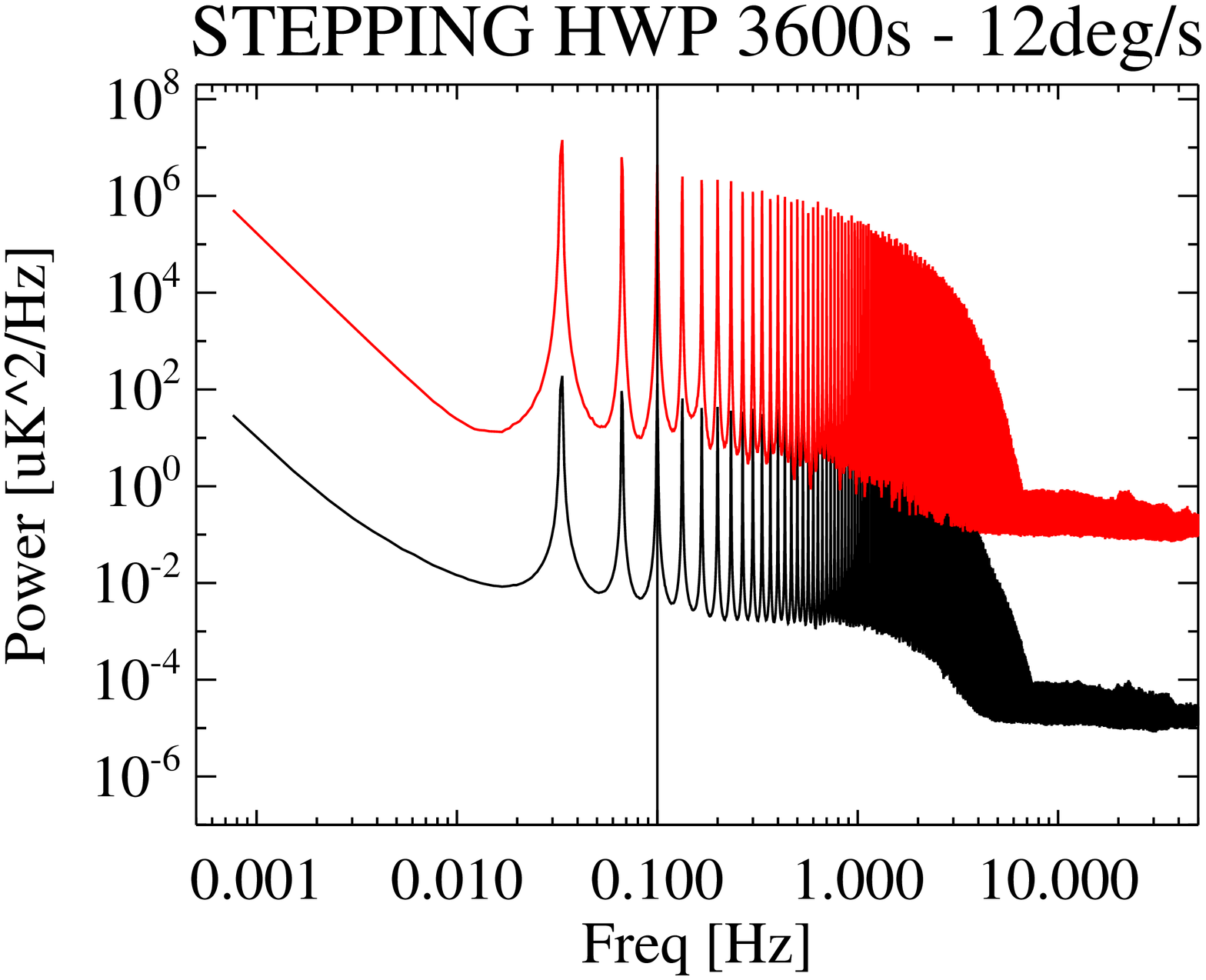}\\
\includegraphics[angle=90,width=0.32\textwidth]{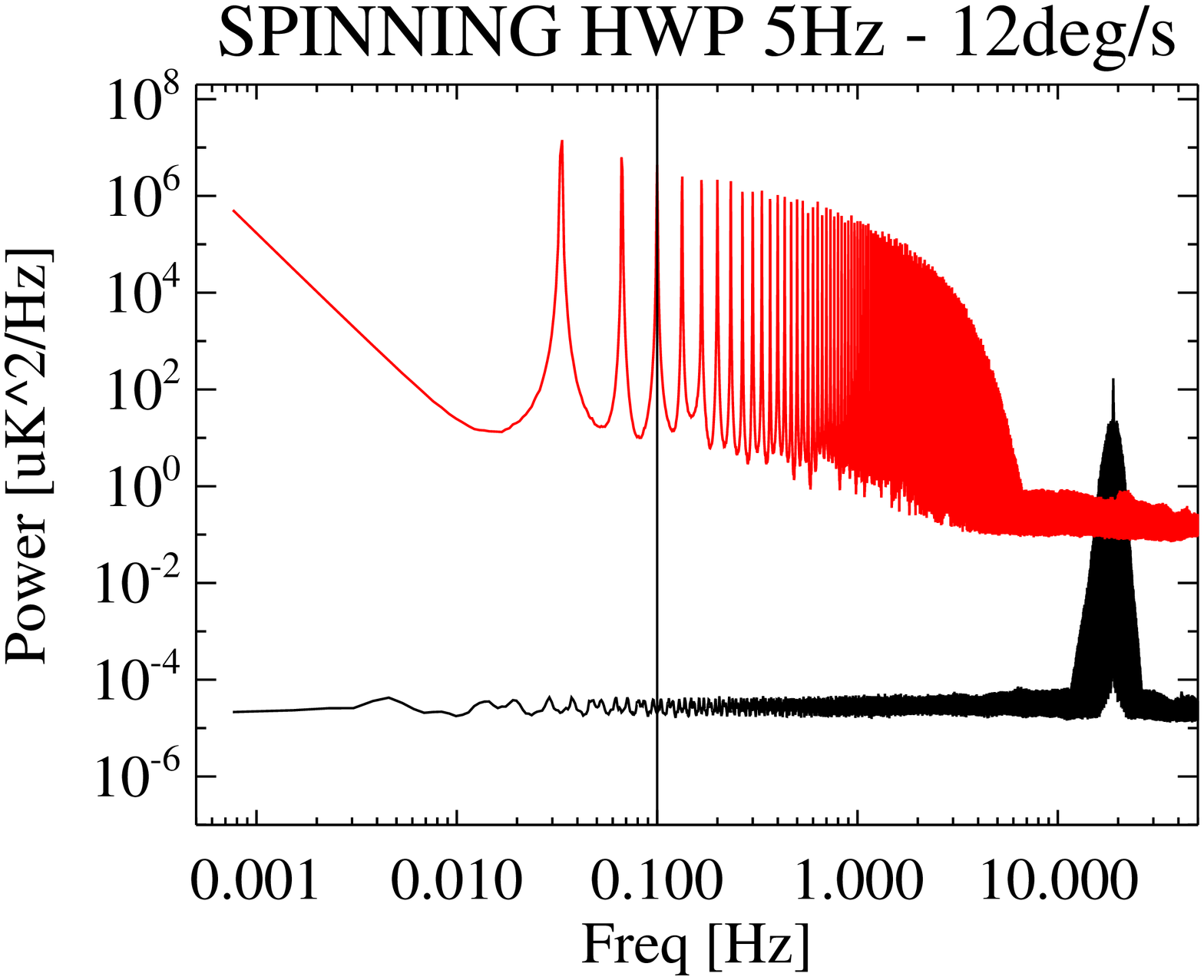}
\includegraphics[angle=90,width=0.32\textwidth]{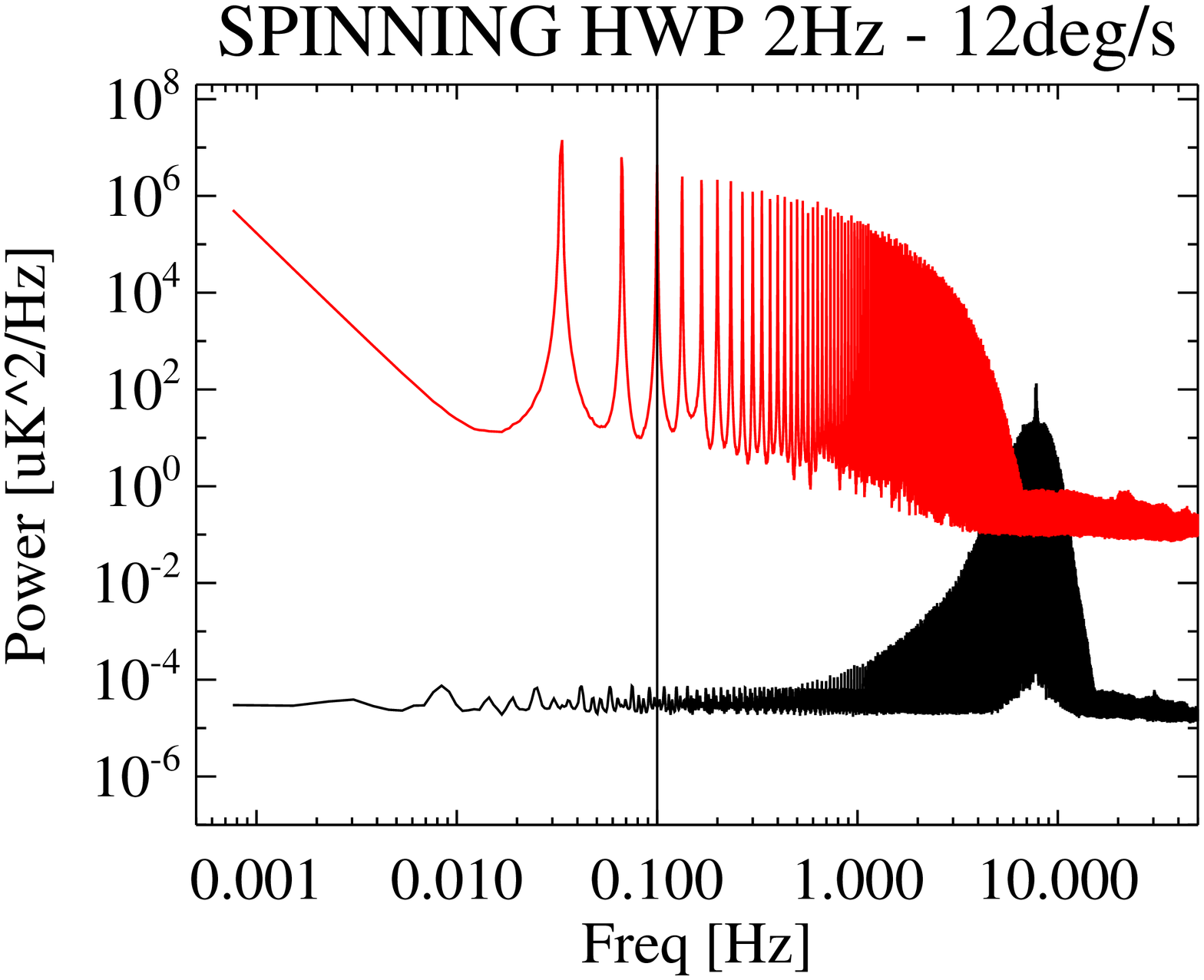}
\includegraphics[angle=90,width=0.32\textwidth]{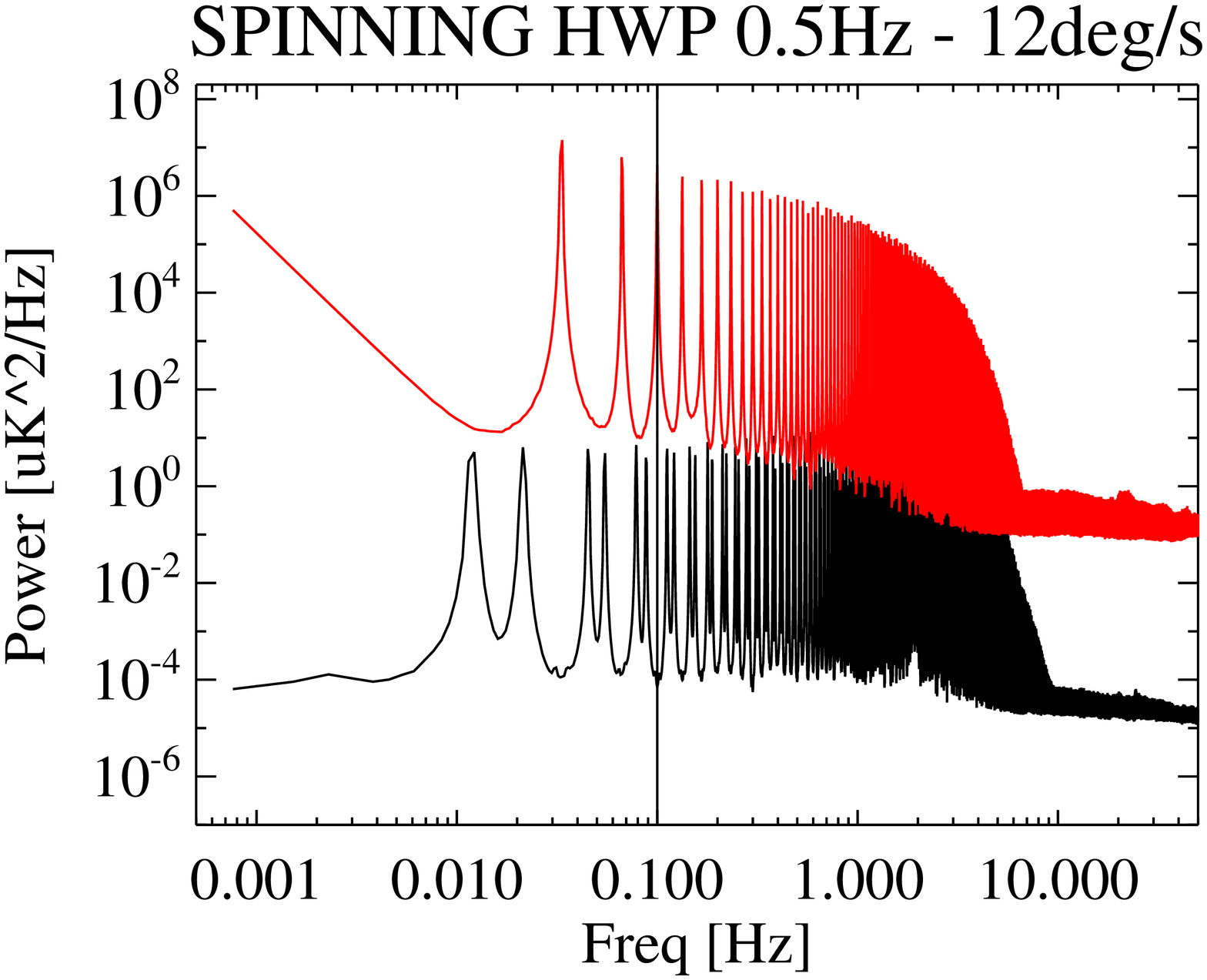}
\caption{Periodograms, i.e. plots of temperature (in red) and polarization (in black) intensity power as 
function of frequency, for the HWP schemes under consideration. On the top, from left to right: stepping HWP 
(1, 60, 3600$\;s$). On the bottom, from left to right: spinning HWP (5, 2, 0.5$\;Hz$).}
\end{center}
\label{period}
\end{figure}

\begin{figure}[H]
\begin{center}
\psfrag{number of pixels}[c][][3]{\normalsize{Number of pixels}}
\psfrag{Pixel inverse cond. number}[c][][3]{\normalsize{Pixel inverse cond. number}}
\includegraphics[angle=90,width=0.4\textwidth]{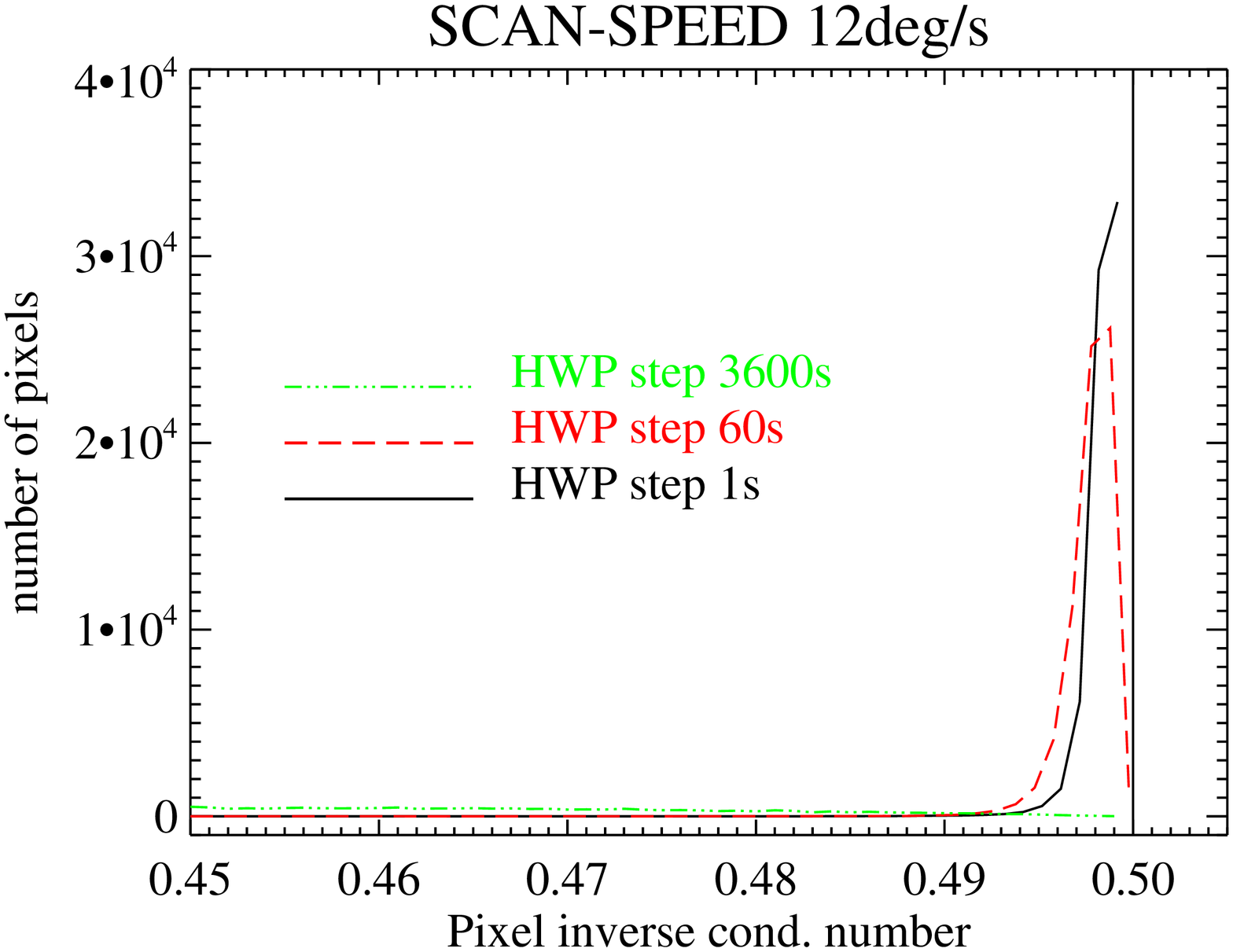}
\includegraphics[angle=90,width=0.4\textwidth]{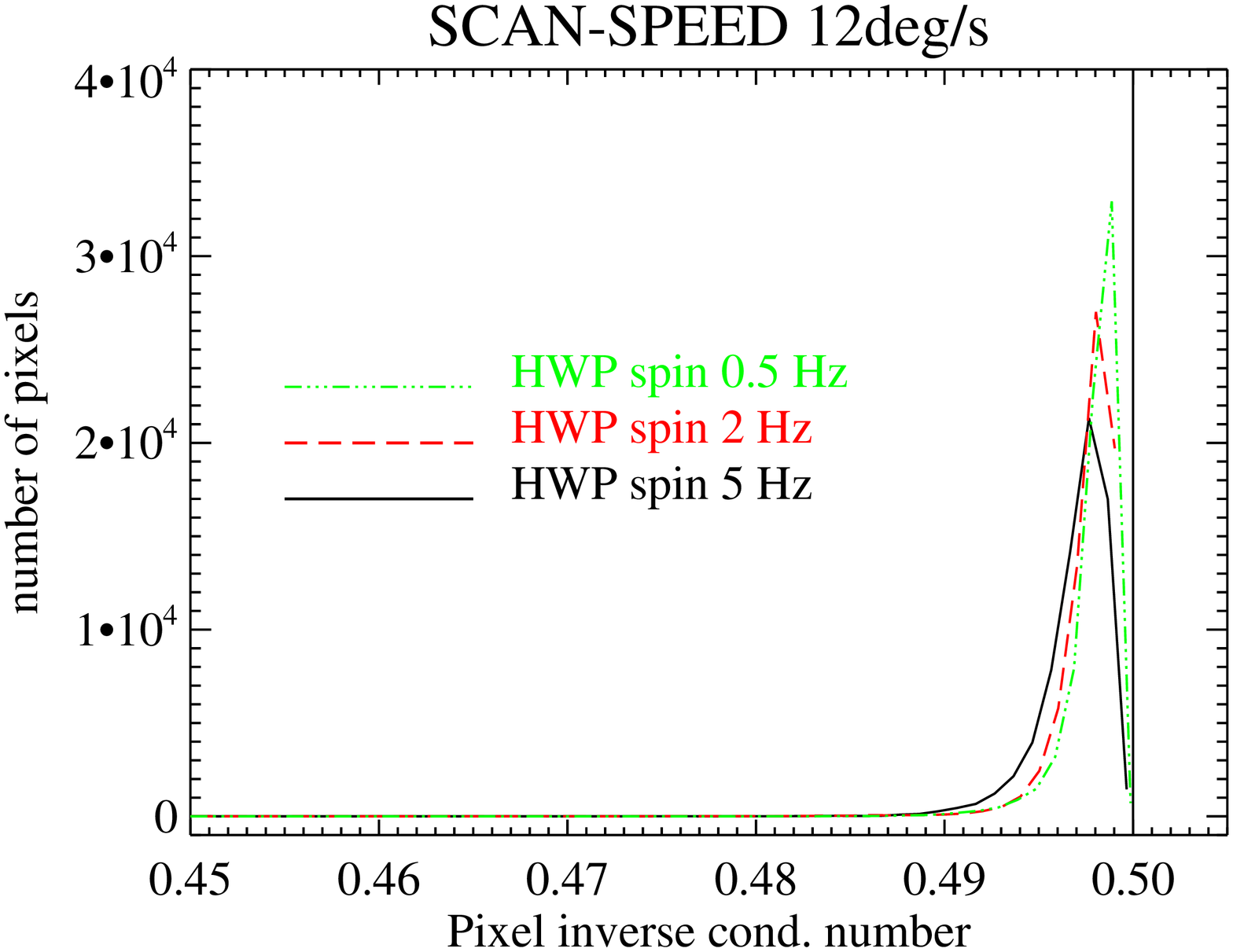}
\caption{Histograms of the pixel inverse condition number for the HWP schemes under exam. 
On the left: stepping HWP (1, 60, $3600\;s$). On the right: spinning HWP (5, 2, $0.5\;Hz$).}
\end{center}
\label{hist}
\end{figure}

In Fig.~\ref{hist} we show the histograms of the pixel inverse condition number, $R_{cond}$, that is an indicator 
of the angle coverage uniformity ($R_{cond} = 0.5$ in the case of perfect pixel coverage). We notice that all the 
HWP configurations under examination provide a very good pixel angle coverage, except the slowest stepping mode 
(at $3600\;s$).

\section{Conclusions} 
A $B$-mode detection would represent an invaluable discovery in cosmology and fundamental physics since it would be 
a ``smoking gun'' evidence for inflation theory. This is due to the specific 
property of the curl polarization component, which is unequivocally sensitive to the tensor mode of primordial 
perturbations as predicted in the inflationary scenario. Among the several experiments designed to observe $B$-modes, 
we focused on the forthcoming LSPE mission. LSPE will consist of a balloon-borne bolometric polarimeter, SWIPE, 
and a ground-based coherent polarimeter, STRIP. The SWIPE instrument will exploit a rotating HWP to mitigate the 
detrimental effects of the instrumental systematics. We tested different HWP configurations, in both stepping and 
spinning modes, keeping fixed the telescope scan-speed. As figures of merit, we analyzed the effectiveness of the 
HWP design in moving the polarization signal to higher frequencies away from the $1/f$ noise and the improvement in 
the pixel angle coverage. We found that the polarization shift to high frequencies is typical of the (fast) spinning 
mode and that any 
HWP scheme provides a very good pixel angle coverage, except a slow stepping mode. We emphasize that our simulations 
do not account for the systematic effects introduced by the HWP itself, such as non-ideal rotation and internal reflections. 
We will face the issue of the 
HWP systematics of its own as well as other figures of merit aimed at the HWP design optimation in a following paper.

\end{document}